\documentclass[journal=ancac3,layout=twocolumn,manuscript=article]{achemso}
\SectionNumbersOn

\makeatletter
\let\l@addto@macro\relax
\makeatother
\usepackage[fontsize=9pt]{scrextend}
\usepackage{titlesec}

\usepackage{charter}

\usepackage{type1cm}
\usepackage{lettrine} 

\renewcommand{\figurename}{\textbf{Fig.}}
\renewcommand{\thefigure}{\textbf{\arabic{figure}}}

\usepackage{amsmath,amssymb}
\usepackage[caption=false,labelformat=parens]{subfig}
\usepackage{graphicx}

\usepackage{geometry}
\usepackage{changepage}
\usepackage{ragged2e}
\usepackage{soul}

\usepackage{caption}
\usepackage[colorinlistoftodos,textsize=small]{todonotes}
\usepackage{bm,bbm}
\usepackage{upgreek,romannum}
\AtBeginDocument{\pagenumbering{arabic}}
\usepackage{dcolumn}
\newcolumntype{.}{D{.}{.}{-1}}
\AtBeginDocument{\usepackage{booktabs}}
\usepackage{color}
\definecolor{darkblue}{rgb}{0.3477, 0.2344, 0.5898}
\definecolor{citeblue}{rgb}{0.2305, 0.4102, 0.6172}
\usepackage{hyperref}
\hypersetup{hyperfootnotes=false,colorlinks=true,citecolor=citeblue,urlcolor=citeblue,linkcolor=black}
\usepackage{memhfixc}

\usepackage{bibunits}
\defaultbibliographystyle{naturemag_noURL}

\makeatletter
\newcommand\footnoteref[1]{\protected@xdef\@thefnmark{\ref{#1}}\@footnotemark}
\makeatother

\newcommand{\GM}{$\overline{\Gamma \text{M}}$}
\newcommand{\GK}{$\overline{\Gamma \text{K}}$}

\let\oldmaketitle\maketitle
\let\maketitle\relax

\author{\small\raggedright\bfseries Anton Tamt\"{o}gl}
\email{tamtoegl@gmail.com}
\affiliation{\small\raggedright Cavendish Laboratory, University of Cambridge, J. J. Thompson Avenue, Cambridge CB3 0HE, United Kingdom}
\altaffiliation{Institute of Experimental Physics, Graz University of Technology, Graz, Austria}
\author{Emanuel Bahn}
\affiliation{\small\raggedright Cavendish Laboratory, University of Cambridge, J. J. Thompson Avenue, Cambridge CB3 0HE, United Kingdom}
\alsoaffiliation{Department of Radiation Oncology, Heidelberg University Hospital, Heidelberg, Germany}
\altaffiliation{Contributed equally to this work}
\author{Marco Sacchi}
\email{m.sacchi@surrey.ac.uk}
\affiliation{Department of Chemistry, University of Cambridge, Lensfield Road, Cambridge CB2 1EW, United Kingdom.}
\alsoaffiliation{Department of Chemistry, University of Surrey, Guildford GU2 7XH, United Kingdom}
\author{Jianding Zhu}
\author{David J. Ward}
\author{Andrew P. Jardine}
\affiliation{\small\raggedright Cavendish Laboratory, University of Cambridge, J. J. Thompson Avenue, Cambridge CB3 0HE, United Kingdom}
\author{Stephen J. Jenkins}
\affiliation{Department of Chemistry, University of Cambridge, Lensfield Road, Cambridge CB2 1EW, United Kingdom.}
\author{Peter Fouquet}
\affiliation{Institut Laue-Langevin, 71 Avenue des Martyrs, 38000 Grenoble, France}
\author{John Ellis}
\author{William Allison}
\affiliation{\small\raggedright Cavendish Laboratory, University of Cambridge, J. J. Thompson Avenue, Cambridge CB3 0HE, United Kingdom}

\date{\today}

\title[Motion of water molecules]{\raggedright\textrm{\huge\bfseries Motion of water monomers reveals a kinetic barrier to ice nucleation on graphene.}}

\begin{document} 

\begin{bibunit}
\twocolumn[
\begin{@twocolumnfalse}
\vspace*{-2cm}
\oldmaketitle

\vspace*{-0.4cm}
{\textcolor{darkblue}{\rule{\textwidth}{1pt}}}

\vspace*{0.2cm}
\textbf{The interfacial behaviour of water remains a central question to fields as diverse as protein folding, friction and ice formation\cite{nicholls1991,rausch2013}. While the structural and dynamical properties of water at interfaces differ strongly from those in the bulk, major gaps in our knowledge at the molecular level still prevent us from understanding these ubiquitous chemical processes. Information concerning the microscopic motion of water comes mostly from computational simulation\cite{tocci2014,ma2015} but the dynamics of molecules, on the atomic scale, is largely unexplored by experiment. Here we present experimental results combined with ab initio calculations to provide a detailed insight into the behaviour of water monomers on a graphene surface. We show that motion occurs by activated hopping on the graphene lattice. The dynamics of water diffusion displays remarkably strong signatures of cooperative behaviour due to repulsive forces between the monomers. The repulsive forces enhance the monomer lifetime ($t_m \approx 3$\,s at $T_S = 125$\,K) in a \textit{free-gas} phase that precedes the nucleation of ice islands and, in turn, provides the opportunity for our experiments to be performed. Our results give a unique molecular perspective of barriers to ice nucleation on material surfaces, providing new routes to understand and potentially control the more general process of ice formation.\\[0cm]
}
{\textcolor{darkblue}{\rule{\textwidth}{1pt}}}
\vspace*{0.2cm}
\end{@twocolumnfalse}
]

\noindent \lettrine[lines=3,nindent=0pt,lraise=0.05]{\textcolor{darkblue}{I}}{ce} often forms easily on solid surfaces and to understand why that happens, the molecular basis of the water-surface interaction needs to be studied\cite{rausch2013,fitzner2014}. The structure, dynamics and chemical properties of water at interfaces differ from those of bulk water and ice\cite{scatena2001,davis2013,kringle2020}. The early stages of ice nucleation involve exceedingly small time and length scales\cite{sosso2016} and while ice nucleation and phase transitions are well understood macroscopically, unravelling the microscopic details presents one of the great challenges in physical sciences with important implications from the chemistry of the Earth's atmosphere\cite{knopf2018} to physicochemical processes occurring on cosmic dust grains\cite{hama2013}. 

It is the motion of water molecules at surfaces, that controls these fundamental phenomena in physics, chemistry and biology as well as a diverse range of technological processes\cite{nicholls1991,rausch2013,tan2020}. Wetting, hydrophobicity and ice nucleation are all very widely studied on the macroscopic scale, using routine methods such as contact angle measurements\cite{kreder2016,schutzius2015,belaeva2020}. However, more precise measurements, with a molecular level of detail, are much scarcer, despite the fact that an understanding could open up new opportunities for the design of advanced materials, by exploiting our ability to tune surfaces at the nanoscale\cite{holst2018}. For example, ice nucleation on surfaces is alone of huge technological relevance to fields as diverse as wind power\cite{kreder2016,parent2011}, aviation\cite{lv2014,schutzius2015} and telecommunications\cite{kreder2016}. 
 
Water is fundamentally challenging to study with atomic resolution. It is difficult to achieve sufficient contrast and resolution with imaging techniques\cite{guo2014}, particularly in order to understand the position of the H atoms and thus the molecular orientation. Electron based techniques such as LEED also scatter weakly from hydrogen and present a severe risk of damage, in the form of water dissociation\cite{hodgson2009,bjorneholm2016}. Some structural studies of water have been possible experimentally, but are usually restricted to flat metal surfaces\cite{hodgson2009,carrasco2012,maier2015,bjorneholm2016,maier2016,shimizu2018} or a few ionic crystals, such as NaCl\cite{guo2014,heidorn2015}. These studies have revealed the role of short range attractive forces. Dynamics and low coverage measurements, which could examine the nature of water interactions more generally, are further complicated by fast diffusion rates and the short lifetimes of water monomers.
Insight has therefore been mostly limited to that possible with numerical simulations \cite{tocci2014,ma2015}, often without any direct experimental validation to support them.

\begin{figure*}[htbp!]
\centering
\includegraphics[width=0.98\linewidth]{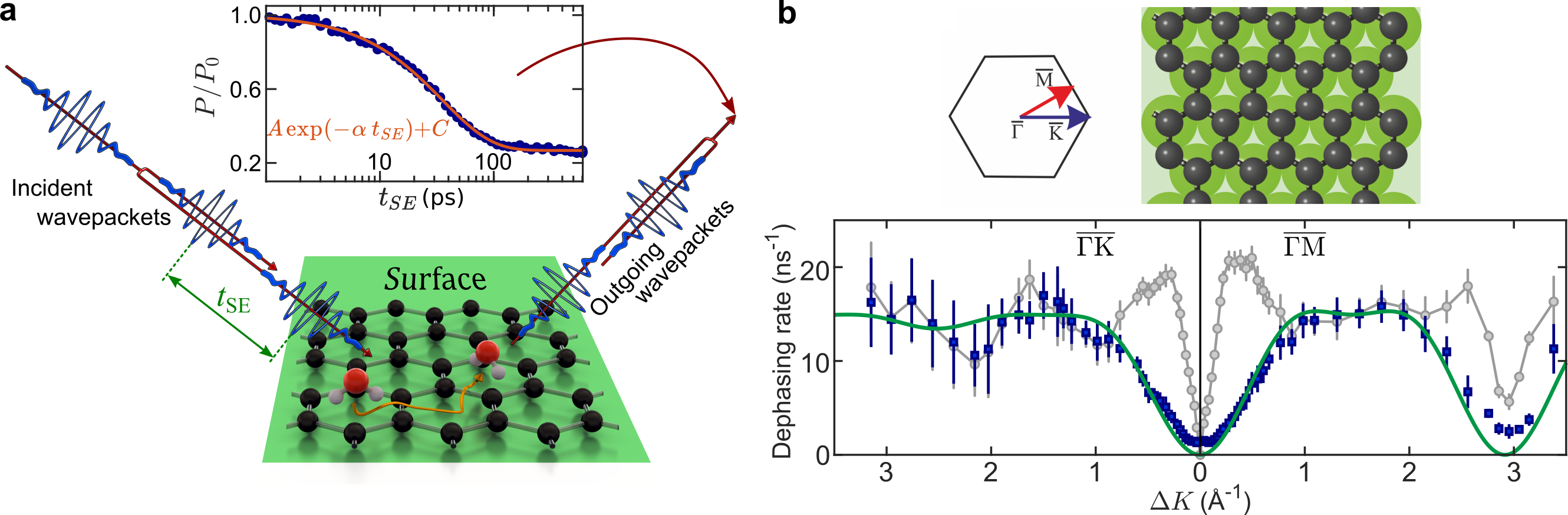}
\caption{\textbf{Diffusion of water monomers on graphene. a} Illustration of the helium spin-echo method: Two wavepackets scatter from the surface with a time difference $t_{SE}$, allowing the motion of molecules on the surface to be interrogated through the loss in correlation, measured through the polarisation of the beam. The inset shows a typical measurement for the diffusion of water on graphene ($T_S = 125$ K, $\Delta K = 0.2$ \AA$^{-1}$). The reduction in surface correlation with increasing spin-echo time follows a single exponential decay (solid line), characterised by the dephasing rate, $\alpha $. \textbf{b} The momentum transfer dependence of the dephasing rate, $\alpha ( \Delta K )$, at $T_S=125$ K from which the mechanism for diffusion follows. Blue data points show single-particle, or incoherent $\alpha ( \Delta K )$, deduced from the coherent scattering data (grey points, see text). An analytical model (green curve) shows the expected behaviour for jumps between the centres of the graphene hexagons. The error bars correspond to the confidence bounds (1$\sigma$) upon determination of $\alpha$ from the measurements - see text.}
\label{fig:Fig1}
\end{figure*}

In this paper we report the serendipitous discovery of a regime where freely mobile water can be studied on a Ni(111) supported graphene surface. We use the helium spin-echo (HeSE) technique, illustrated in \autoref{fig:Fig1}a, to measure surface correlations in the water monomer motion (see \nameref{methods}). HeSE uses wavepacket splitting and recombination to give temporal sensitivity over picosecond timescales, resulting in data of the form shown in the inset of \autoref{fig:Fig1}a. The use of these very low energy He atoms completely excludes any possibility of damage or dissociation of the water. As we describe below, by analysing the dephasing rates in the correlation measurements to obtain the signatures shown in \autoref{fig:Fig1}b, we are able to establish that, contrary to expectations, strong repulsive interactions exist between adsorbed water molecules.  We attribute these forces to dipolar interactions, arising from structural hindrance of water reorientation by the adsorption geometry. The repulsion leads to an effective kinetic barrier to ice formation, which both makes our measurements possible, and suggests new intelligent strategies are possible for controlling and directing the nucleation and growth of ice.

\subsection*{Results}
\begin{figure}[htbp!]
\centering
\includegraphics[width=0.8\linewidth]{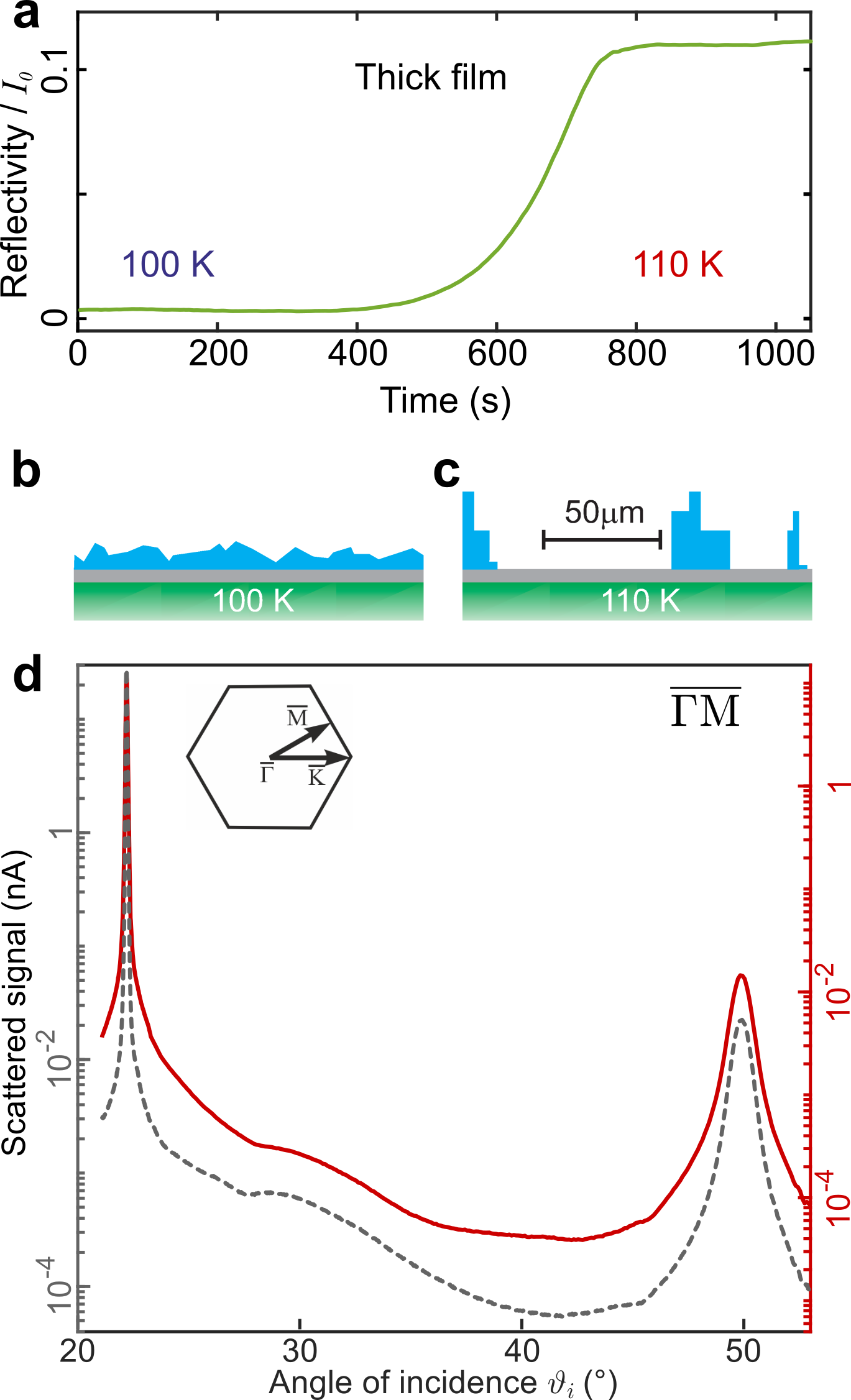}
\caption{\textbf{De-wetting of a thick film and island formation} \textbf{a} A thick water film prepared at 100\,K has low helium reflectivity, indicating a rough surface (time $< 200$\,s). The temperature is increased to 110\,K, between 100-300\, s, and the reflectivity rises later, as the graphene substrate is revealed (time $> 600$\,s) - see main text. \textbf{b} and \textbf{c} are cartoons indicating the morphology of the surface at 100\,K and 110\,K respectively. In both cases it is probably amorphous solid water that forms. At 110\,K de-wetting of the surface occurs to reveal pristine graphene. The scale bar gives a qualitative indication of the island separation (see \nameref{S:Coverage}). \textbf{d} De-wetting is confirmed as helium-diffraction from the graphene film, before water adsorption (grey dashed curve), is identical to that from the thick water film after heating to 110\,K (red curve). Both curves are measured at 110\,K with an incident beam energy, $E_i = 8 ~\mbox{meV}$, plotted with different ordinate limits on the left and right-hand side. The hexagon shows the principle symmetry directions of the surface Brillouin zone.}
\label{fig:Fig2}
\end{figure}

In order to identify the range of conditions where individual water molecules are mobile, we carried out extensive adsorption and desorption measurements on the graphene/Ni(111) surface. The substrate was prepared in ultra-high vacuum (UHV) conditions and graphene was grown using established methods\cite{tamtogl2015} (see \nameref{methods} and \nameref{S:prep} in the supplementary information (SI)). Growing a thick film of water at 100\,K results in a very low helium reflectivity, which is typical of disordered structures\cite{yang2009,andersson2007,souda2018} as confirmed by the lack of any helium diffraction. The data indicate an amorphous solid water layer covering the entire surface. Heating the surface slightly leads to a significant change. \autoref{fig:Fig2}a shows how the reflectivity increases over a period of minutes at 110\,K. We can rule out desorption as direct measurements show, desorption is negligible at this temperature (see \nameref{S:prep}). Simultaneously, diffraction peaks emerge at the positions of a graphene lattice, as shown by the red curve in \ref{fig:Fig2}d. The relative diffraction intensities are identical to the pristine graphene surface (grey dashed-curve), which would not be the case for a crystalline ice overlayer\cite{corem2013}. Ice \Romannum{1}\textsubscript{h} and ice \Romannum{1}\textsubscript{c} also have too large a lattice spacing\cite{kimmel2009} to give rise to this periodicity, even for the spacing in the recently discovered square ice \cite{algara2015,kimmel2009}.  The diffraction pattern indicates that large areas of graphene are exposed, alongside localised areas with multi-layer ice islands on the surface.  Thus, we conclude that the deposited water has migrated to form isolated islands of amorphous ice, as illustrated in \autoref{fig:Fig2}c. Such behaviour is consistent with the strongly hydrophobic behaviour previously seen on pristine graphene\cite{shih2013,zhao2017,belaeva2020} and a similar behaviour has been observed for water on other metal-supported graphene systems\cite{standop2015,kimmel2009}. The formation of islands provides the first indication that in this regime, water molecules must be able to diffuse freely over the graphene surface.

\begin{figure}[htbp!]
\centering
\includegraphics[width=0.8\linewidth]{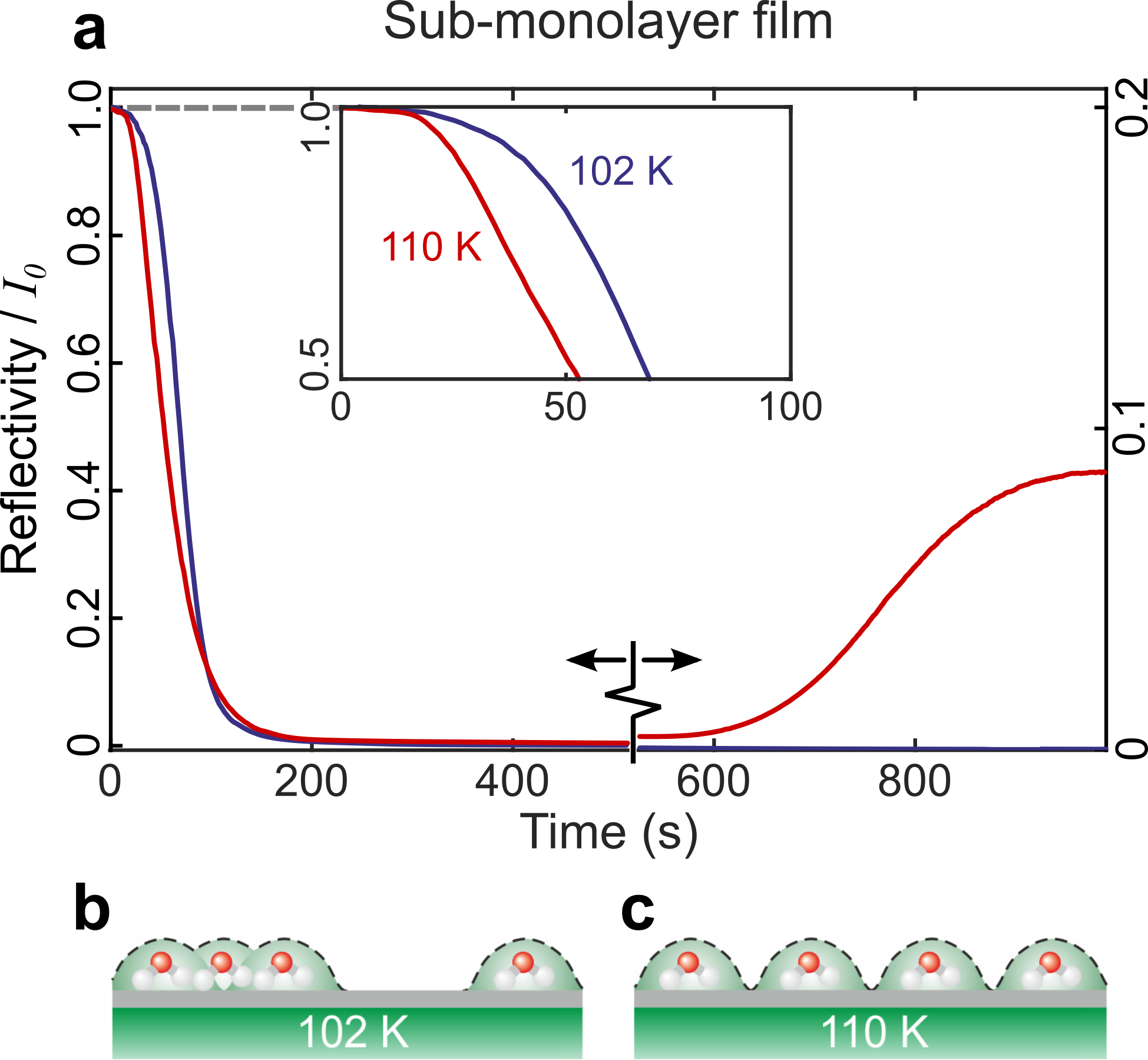}
\caption{\textbf{Molecular adsorption in the sub-monolayer regime.} Helium reflectivity from a sub-monolayer film, shown as a function of time during constant deposition of water molecules. The reflectivity is a measure of the fraction of the original, pristine surface that is exposed. At low coverages (time $< 100$\,s), the intensity drops as individual water molecules obscure the underlying graphene. The blue curve shows adsorption at 102\,K, a temperature where the morphology of thick film measurements (\autoref{fig:Fig2}) suggest the molecules are immobile on the time-scale of the experiment. \textbf{b} Cartoon illustrating the random distribution of immobile molecules. The reflectivity falls rapidly due to the large scattering cross-section of each molecule (dotted lines) despite considerable overlap, even at a fraction of a monolayer. The reflectivity remains low as the thick film, shown in \autoref{fig:Fig2}b, is formed.  The red curve in panel \textbf{a} shows the same rate of adsorption at $110~$K, where the reflectivity drops even more quickly (time $< 100$\,s). The faster drop is due to isolated water molecules obscuring a greater area of the graphene surface, as shown in \textbf{c}. We attribute the difference to an increased mobility allowing the molecules to achieve a lower energy configuration. It is the first indication that repulsive interactions must keep the water molecules separated. At longer times and higher coverage (time $> 600$\,s), the mobile molecules eventually nucleate ice structures, as discussed in the main text. The surface de-wets, allowing scattering from the underlying graphene to re-emerge, exactly as shown in \autoref{fig:Fig2}c.}
\label{fig:Fig3}
\end{figure}

The observation of de-wetting and island formation is supported by helium reflectivity measurements at very low coverage.  \autoref{fig:Fig3}a compares the behaviour at a surface temperature of 102\,K (blue curve), where the molecules are immobile and at 110\,K (red curve), where there is some mobility. In both cases, the reflectivity falls sharply when water molecules begin to adsorb because of their large cross-section for diffuse scattering.  At 102\,K (blue curve), the reflectivity drops to near zero, and remains there throughout the experiment. At 110\,K (red curve), the initial drop is even more rapid, despite the same rate of water uptake onto the surface. The observation can be understood if water molecules stay further apart at 110\,K than 102\,K, as shown schematically in \autoref{fig:Fig3}b and 3c, where the large scattering cross section of each molecule is indicated by the dotted lines. Regularly spaced molecules at 110\,K reduce the overall cross section overlap, and thus result in the faster reduction in reflectivity with coverage. The additional spacing must arise from water mobility at 110\,K, which we discuss below. After further exposure at 110\,K, the reflectivity recovers due to the nucleation of islands, at a rate much slower than the molecular diffusion \cite{leenaerts2009,andersson2007}. Water molecules desorb at higher temperatures, but up to approximately 130\,K it is possible to maintain a constant coverage by applying an over-pressure of water. Under these equilibrium conditions, water monomers diffuse continuously between islands of ice, producing an elusive ``free-gas'' of monomers, which allows us to study the interactions between isolated molecules prior to ice formation. Differences that might be expected in terms of the film growth with respect to different metal substrates and bulk graphite\cite{andersson2007,smith2014,souda2018,kringle2020} are further outlined in the \nameref{S:Discussion}.

In the window of dynamical equilibrium between 113\,K and 130\,K, HeSE experiments were performed to make detailed measurements of water molecular motion. Temporal correlation functions of the form illustrated in the top inset of \autoref{fig:Fig1}a describe the  dephasing in the scattering due to the motion of water over picosecond time-scales and on reciprocal space length-scales between 2\,\AA~and 200\,\AA. Each measurement can be described by a single exponential decay function, $A \exp (-\alpha \, t_{SE}) + C$, as illustrated by the red line, to obtain the rate of dephasing, $\alpha$. (Uncertainties are the corresponding confidence bounds (1$\sigma$) of the single-exponential fit.) The variation of $\alpha$ in reciprocal space, i.e. as a function of the surface parallel scattering momentum transfer $\Delta K = | \Delta \mathbf{K} |$, provides a signature of both the mechanism and rate of the molecular motion.

The grey points in \autoref{fig:Fig1}(d) show the variation of $\alpha ( \Delta K )$ for water motion at 125\,K along two directions in reciprocal space, at a relative coverage of 0.07 monolayers (see \nameref{S:Coverage}). Helium atoms scatter coherently and the effects of correlations due to long-range forces between the monomers play a role in the scattering, especially at values of $\Delta K <$ 1\,\AA.  We discuss the full picture later but first we identify the energy-landscape for motion by removing the effects of pair-correlations among the adsorbates using established methods\cite{pusey1975,leitner2009,ward2020}. We use an approximate form for the scattering form factor\cite{ward2020} and estimates of the quasi-elastic structure-factor (see \nameref{S:MCSupp}) to obtain the result for incoherent scattering and the corresponding single-particle dephasing-rate \autoref{fig:Fig1}b (blue points). 

The single-particle dephasing-rate (blue points) is periodic in $\Delta K$ rising from the origin and returning to $\alpha=0$ at about $\Delta K = 2.9$ \AA$^{-1}$ in the \GM\ direction. The periodicity indicates that motion takes place by a jump mechanism and the data is well described by a simple model for jump diffusion\cite{jardine2009},
\begin{equation}
\alpha ( \Delta K ) =\frac 2{\tau } \, \sum _mp_m\cdot \sin ^2\left(\frac{ \Delta \mathbf{K}\cdot \mathbf{j}_m }
2\right),
\label{eq:CE}
\end{equation}
where the particle rests for a time $\tau$ within an adsorption site on the corrugated surface, before moving instantaneously to another equivalent adsorption site in a neighbouring unit cell along the vector $\mathbf{j}_m$, with probability $p_m$.

The analytic form is shown as a green curve in \autoref{fig:Fig1}b and corresponds to jumps on the hexagonal graphene lattice, where the jump length is equal to the lattice constant and multiples thereof. The curve includes jumps to nearest, next-nearest, and second-nearest neighbours, giving a residence time of $\tau  = 65 \pm 3\,\mbox{ps}$ and a relative jump contributions of $p_n = 63$ \%, $p_{nn} = 20$ \%, and $p_{nnn}$ = 17 \%, respectively. Importantly, a jump model can only describe the experimental data if the water molecule is adsorbed in the centre of the hexagons formed by the carbon rings. Jumps with other adsorption geometries would either give rise to a different dependence upon $\Delta K$ or to the appearance of multiple exponential decays in the data\cite{tuddenham2010}. Hence we can unambiguously determine the adsorption site of water on graphene. These findings are in good agreement with our DFT calculations and with reported angle-resolved photoelectron spectroscopy of H$_2$O on graphene/Ni(111) which have been interpreted in terms of a preferential adsorption on either hollow or bridge sites\cite{bottcher2011}.
\begin{figure}[htbp!]
\centering
\includegraphics[width=0.8\linewidth]{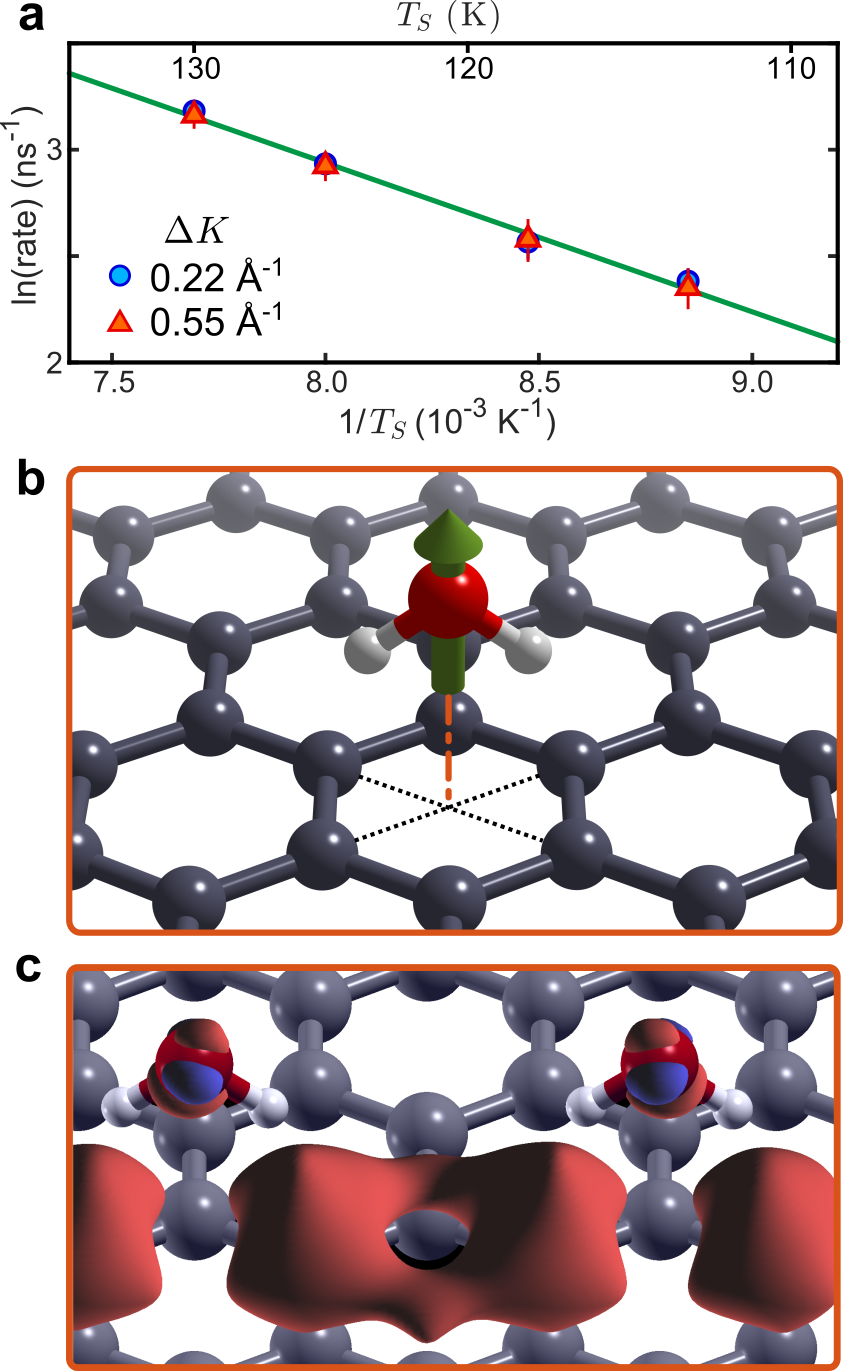}
\caption{\textbf{Temperature dependence and theoretical results for water adsorbed on graphene. a} The temperature dependence of $\alpha$ can be used to determine the activation energy for diffusion of water on graphene. A constant surface coverage of $0.07$ ML, corresponding to a reflectivity attenuation factor of 4, was maintained at all temperatures by adjusting the over-pressure applied. \textbf{b} Adsorption geometry of water with the green arrow illustrating the direction of the net dipole of the water molecule. \textbf{c} Charge density difference for two water molecules adsorbed on graphene (red/blue isosurfaces correspond to $\pm 0.0025\,\mbox{e} / \mbox{\AA}^3$) illustrating the dipole moment. The dipole moment of a water monomer on graphene is $6.4\cdot 10^{-30}~\mbox{C}\mbox{m}=1.9\,\mbox{D}$, which is slightly larger than for an isolated water molecule.}
\label{fig:Fig4}
\end{figure}

An accurate tracer diffusion coefficient, $D$, for two-dimensional water motion can be calculated from the residence time, $\tau$, using,
\begin{equation}
 D=\frac 1{4\tau }\bigl\langle l\bigr\rangle ^2,
\label{eq:diff}
\end{equation}
where $\langle l \rangle = 3.3\,\mbox{\AA}$ is the average jump length.  Based on the result for single particle motion, we obtain a diffusion constant $D = (4.1 \pm 0.2) \cdot 10^{-10}\,\mbox{m}^2/\mbox{s}$ at 125\,K. An activation energy, $E_a$, for motion on a particular length-scale can be obtained from temperature dependent measurements using Arrhenius' law,
\begin{equation}
\alpha =\alpha _0\cdot \exp \left\{  - E_a / \left( k_B\cdot T_S \right) \right\} \; ,
\label{eq:Arrhenius}
\end{equation}
where $\alpha_0 $ is a pre-exponential factor relating to the jump attempt frequency, $k_B$ is the Boltzmann constant and $T_S$ is the surface temperature. \autoref{fig:Fig4}a plots $\ln (\alpha )$ versus $1/T_S$ and shows a clear linear dependence, giving $E_a = (60 \pm 4)\,\mbox{meV}$ and $\alpha_0 = (5 \pm 1 )\,\mbox{ps}^{-1}$. (Obtained from a weighted linear fit with confidence bounds of 1$\sigma$.) There is very little difference between the values obtained at the two different momentum transfers shown in \autoref{fig:Fig4}a. Our diffusion rates are significantly lower than those from recent molecular dynamics (MD) simulations\cite{tocci2014,ma2015}, which show very fast diffusion of water droplets and estimated a diffusion coefficient of $6 \cdot 10^{-9}~\mbox{m}^2/\mbox{s}$ for single water molecules at 100\,K\cite{ma2011}. We note those calculations were performed on free standing graphene while our measurements are on Ni(111) supported graphene and in particular, the ripples giving rise to the ultra-fast droplet diffusion\cite{ma2015} are suppressed by the substrate\cite{tamtogl2015}. We also note that our measurements on graphene indicate a higher diffusion and hopping rate than experimental values for other substrates\cite{bertram2019,tamtogl2020,heidorn2015} (see Supplementary Table \ref{tab:CompTab} in \nameref{S:Discussion}).

We now turn to the interactions between water molecules, which are encoded in the differences between the coherent and incoherent rates in \autoref{fig:Fig1}b (blue and grey data points). A series of kinetic Monte Carlo (KMC) simulations were performed to determine the nature of these forces. The hexagonal hopping model described earlier was combined with an interaction potential, $V_{pp}$, between water molecules, of the form
\begin{equation}
V_{pp} = \pm\frac{B}{r^3}=\pm\frac{p^2}{4 \pi \epsilon_0 r^3},
\label{eq:MCdipole}
\end{equation}
which represents a pairwise dipole-dipole interaction, where $p$ is the effective value of the dipole moment and $r$ is the distance separating the two dipoles. Allowing for either positive or negative prefactors provides the ability to explore both repulsive and attractive interactions. Using trajectories from the KMC simulation, simulated coherence functions and dephasing rates were obtained to compare with the $\Delta K$ resolved experimental data, while also adjusting the simulation to reproduce the temperature dependent measurements.
\begin{figure}[htbp!]
\centering
\includegraphics[width=\linewidth]{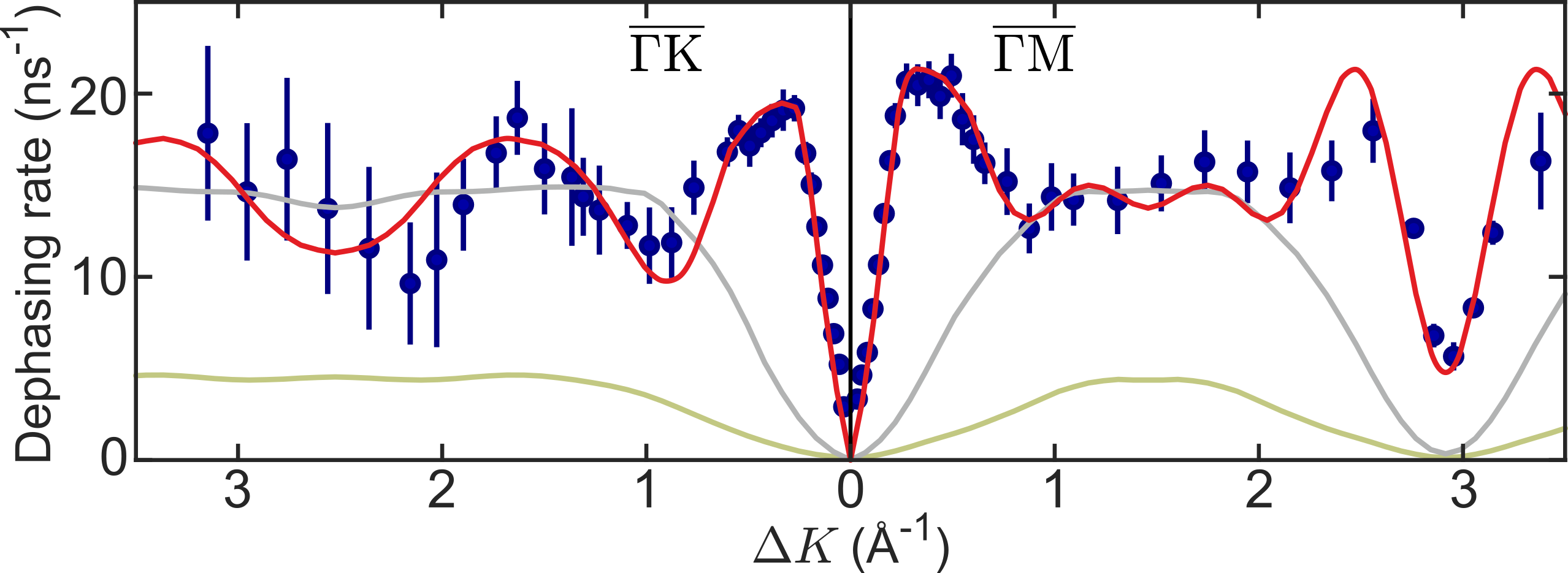}
\caption{\textbf{Evidence for repulsive interactions in the diffusion process.} Experimental dephasing rates for coherent scattering (blue dots) compared with kinetic Monte-Carlo calculations (solid curves). The calculations add a force (\autoref{eq:MCdipole}) to the hopping model derived earlier.  The experimental data are described well by repulsive dipole forces (red curve) but models using attractive forces (green curve) or no forces (grey curve) cannot reproduce the data.  Note that the model without forces (grey curve) is, as expected, similar to the analytic curve for incoherent scattering shown in \autoref{fig:Fig1}.}
\label{fig:Fig5}
\end{figure}

\autoref{fig:Fig5} compares the experimental data with KMC simulation results. The red line shows the case of repulsive interactions, where the dipole moment of each molecule has been adjusted to best fit the experimental data, giving a value of $p=1.8\pm 0.2$\,D. There is excellent agreement with the measurements, particularly around the steep rise at $0.5\,\mbox{\AA}^{-1}$, a characteristic feature of adsorbate interactions, and around the minimum at $2.9~\mbox{\AA}^{-1}$ in the $\overline{\Gamma} \overline{\textrm{M}}$ direction, where the same phenomenon is seen due to periodicity. The magnitude of the dipole moment obtained from the experiment is in good agreement with the value of $p=1.9$\,D obtained from our DFT calculations. Attractive interactions, by contrast, cannot describe the data. The green curve in \autoref{fig:Fig5} replaces repulsion with attraction of the same strength.  Attractive forces suppress the jump rate and do not reproduce the correct dependence on $\Delta K$.  The reduction in rate is progressive as the magnitude of the forces increases, as might be expected. When there are no interactions (\autoref{fig:Fig5}, grey line) the particles move independently and we obtain the same form as the analytical model for hopping in Figure 1(b).

Together, these observations provide conclusive confirmation of significant long-range {\em repulsive} interactions between water monomers. These interactions act to keep the water monomers apart, thus suppressing nucleation and increasing both monomer lifetime and mobility. It is well known that short range attractive forces and hydrogen bonding play a major role in the clustering of water.  We therefore conclude that there must be an intermediate barrier between the long-range repulsive regime and the short range attractive one. This intermediate barrier must be overcome for ice nucleation to occur.  Note that this barrier is in addition to the activation energy required to hop between sites and the energy required to reorient the molecules relative to each other. The additional energy barrier we observe represents a fundamental change in our understanding of ice formation on surfaces, where the intermolecular forces have been assumed to be exclusively attractive.

Questions about the origin and form of the new barrier immediately arise and how they relate to the dipole moment of the water molecules\cite{guinea2016}. Our DFT calculations, shown in \autoref{fig:Fig4}b-c, indicate that water molecules all adsorb with the same orientation, with a dipole moment slightly larger than for an isolated water molecule. The alignment of individual dipoles perpendicular to the surface plane, leads to strong repulsive interactions and is the likely origin of the forces we observe. In order for a water cluster to nucleate, molecules must first come into close proximity and must then re-orient to adopt a hydrogen bonded configuration. Both of these steps have separate energy barriers, such that in combination they strongly inhibit the overall process. Repulsive interactions between adsorbates occur widely at surfaces and can limit the density of adsorbed species as well as defining the adsorbate structure (see for example\cite{lischka2002, sendner2007,lukas2001,torrente2007,yokoyama2007, stadler2009}). In the present work, a kinetic barrier arising from repulsive forces of adequate range is a different mechanism that provides a new insight into the inhibition of ice nucleation.

Cluster nucleation is closely related to the process of water molecule attachment to existing islands -- although once ice has nucleated, dipole repulsion from the island is reduced, such that only the re-orientation barrier is relevant. The lifetime of each water molecule on the surface before it sticks to an island can be estimated from the ratio of the equilibrium coverage, $0.07\,\mbox{ML}= 8\cdot 10^{17}\,\mbox{m}^{-2}$, and the adsorption rate, $2.5 \cdot 10^{17}\,\mbox{m}^{-2}\mbox{s}^{-1}$, giving roughly 3\,s at 125\,K (see \nameref{S:Coverage}). The very long lifetime compared to the residence time in any given adsorption site reflects the difficulty in overcoming any one of these barriers. We also suggest that desorption of water molecules in our experiments is only likely to happen from the surface of water islands since our experimentally determined desorption energy of $(520\,\pm 20)\,\mbox{meV}$ (see \nameref{S:TDS}) is close to the sublimation enthalpy from ice\cite{chakarov1995,bolina2005}.

\subsection*{Discussion}
Finally, we can consider our results in the broader context of ice formation. Our measurements on water monomers were only possible as we discovered a regime where, within a small temperature window, individual water molecules diffuse in dynamic equilibrium with islands of ice and where the molecules have life-times long enough to apply HeSE measurements. The lack of any other data on water monomer dynamics means it is difficult for us to make completely definitive statements about the generality of our observations. However, dipole formation occurs widely upon adsorption\cite{mainak2017} and whenever those dipoles are prevented from re-orienting during diffusion, dipole-driven inter-adsorbate repulsion is viable. Such interactions have been observed between hydrocarbons\cite{lukas2001} and between alkali metals\cite{alexandrowicz2006} on metal substrates, but to our knowledge the present work is the first report of a powerful suppression of nucleation that arises from strong inter-molecular repulsion between water molecules. Similarly, the significance of repulsion in the context of suppressing nucleation has not previously been recognised. Attractive forces and hydrogen bonding, which are dominant after the onset of ice growth at higher coverage, have always been assumed. It represents an important step in unravelling the unique behaviour of ice and the complex relationships between adsorption, jump diffusion and long-range inter-molecular interactions. Our new understanding also suggests new, broadly applicable, strategies for further suppressing or otherwise controlling the ice nucleation process, by enhancing the dipole formed during adsorption. Such an effect could be achieved by, for example, using surface treatments leading to greater electron transfer, or in the case of graphene by altering the supporting substrate. In these respects, the hydrophobic character of the graphene substrate\cite{shih2013,zhao2017,an2017} and particularly the adsorption geometry play important roles, but it seems reasonable to expect the dipolar effect could apply much more generally. The fundamentally new understanding of water transport and assembly on the nanoscale that emerges from our work has the potential to become a new paradigm.  We hope that the wide ranging implications for fields as diverse as nanofluidics, astrophysics and biology will stimulate a wide range of new research, understanding and application.

\subsection*{Methods}
\label{methods}
\paragraph{Experiment and sample preparation}
Gaining direct images of water on non-metallic surfaces remains challenging because of the weak interaction of single water molecules with those substrates. E.g. on graphene, water has previously only been visualised when subsurface, due to its dynamic nature\cite{xu2010,he2012}. Compared with other techniques, He atom scattering (HAS) has the advantage of being the most delicate surface-probing technique and is sensitive to H atoms in the top layer\cite{corem2013,avidor2016,bahn2016,lin2018}. All measurements have therefore been performed using the Cambridge helium-3 spin-echo facility (HeSE)\cite{jardine2009,jones2016}. The schematic principle of He spin-echo is illustrated in \autoref{fig:Fig1}a (main text): A polarised He beam, illustrated by the blue wavepacket, is split into two components which are separated in time by $t_{SE}$. After scattering from the surface, the separated wavepackets are recombined. If the surface changes between scattering of the two parts of the wavepacket, a loss of polarisation is observed in the detected beam, which is directly related to the change in surface correlation and in the case of surface diffusion, usually follows an exponentially decaying form (see Jardine \textit{et al.}\cite{jardine2009} for more information).

The preparation of a single graphene layer on Ni(111) is described in Ref.\cite{tamtogl2015} and the supplementary information (\nameref{S:prep}). Water was dosed onto graphene with a microcapillary array beam doser which was brought close to the surface. During H$_2$O dosing, the partial pressure of water in the scattering chamber was maintained using an automatic leak valve, and the helium reflectivity monitored. During dynamics measurements, dosing was adjusted to achieve a certain attenuation of the helium reflectivity, which corresponded to a particular coverage. Reflectivity was regularly checked to ensure equilibrium was maintained during individual experiments and between measurements under the same conditions, to ensure reproducibility.

A microcapillary array beam doser was used for depositing water on the nickel surface.  The doser was moved to $5$ cm distance from the sample to reduce the water load in the scattering chamber and the dose was estimated from the water pressure in the chamber and the enhancement factor, which is known from previous work\cite{ward2013}. Water was supplied to the doser from a baked stainless steel tube filled with de-ionised water, using the vapour pressure over the liquid phase at room temperature.  The water was purified using a process of several freeze-pump-thaw cycles, where the water inside the tube was frozen and the gas phase above the frozen ice was pumped away. Several repeated cycles were performed until a quadrupole mass spectrometer in the scattering chamber only showed pure water. The water was re-purified prior to every series of adsorption, diffraction, or He spin-echo measurements and regular mass spectrometer scans were performed throughout the measurements to exclude the possibility of contamination.

\paragraph{DFT calculations\label{DFT}} We performed calculations using CASTEP\cite{clark2009}, a plane wave periodic boundary condition code. The Perdew Burke Ernzerhof\cite{perdew1996} exchange correlation functional, with the dispersion force corrections developed by Tkatchenko and Scheffler (TS method)\cite{tkatchenko2009}, was employed for all the calculations presented in this work. The plane wave basis set was truncated to a kinetic energy cutoff of 360 eV. The calculations are performed on a $(6\times 6)$ graphene cell, carbon atoms are fixed, $k$-point sampling has been done with a $(2\times 2 \times 1)$ MP grid\cite{monkhorst1976}. A vacuum layer of $15~\mbox{\AA}$ was imposed above the graphene surface in order to avoid spurious interactions with the periodically repeated supercells. All the calculations use Vanderbilt Ultrasoft Pseudopotentials\cite{vanderbilt1990} and the $x$, $y$ coordinates of the O-atoms are fixed. The electron energy was converged up to a tolerance of $1\cdot 10^{-8}~\mbox{eV}$ while the force tolerance for the geometrical optimisations was $0.05~\mbox{eV} / \mbox{\AA}$.

\paragraph{Kinetic Monte-Carlo simulations} Kinetic Monte Carlo (KMC) simulations employing a modified form of the Metropolis algorithm were used to provide insight into the mechanism of adsorbate interactions during diffusion\cite{nissila2002,tamtogl2020,sakong2020}. Water molecules move on a hexagonal lattice with jumps up to third nearest neighbour sites. A periodic $(60\times 40)$ grid was used, where H$_2$O molecules were were initially located on grid sites at random. The potential energy for a molecule at each site in the grid was calculated for the initial configuration, taking into account repulsive/attractive inter-adsorbate interactions using a pairwise dipole-dipole potential of the form described in \autoref{eq:MCdipole} (main text).

Each MC step consists of choosing a water molecule at random which may then hop to one of its neighbouring sites, with specific probabilities for jumps to first, second and third nearest neighbours. Provided that the water molecule is not blocked from entering the new site by another molecule, the probabilities are weighted by the difference in the potential of the molecule at the two sites. If several new sites with lower potential energy exist, one of them is chosen at random and the molecule is moved into the new site.

\subsection*{Data and code availability}
\paragraph{Data availability} The data supporting the findings of this study are available from \url{https://doi.org/10.17863/CAM.55076}.
\paragraph{Code availability} The code for the kinetic Monte Carlo simulations is available from \url{https://doi.org/10.5281/zenodo.3240428} under the GNU/GPL-3.0 license.

\renewcommand{\refname}{\large References}
\setlength{\bibsep}{0.0pt}
\putbib[literature]
\end{bibunit}

\subsection*{Acknowledgements} 
The authors would like to thank G. Alexandrowicz for many helpful discussions. A. Tamt\"ogl acknowledges financial support provided by the FWF (Austrian Science Fund) within the project J3479-N20. M. Sacchi would like to acknowledge the Royal Society for funding his research through the University Research Fellowship {URF/R/191029} and the UK's HEC Materials Chemistry Consortium, which is funded by EPSRC (EP/R029431), for time on the ARCHER UK National Supercomputing Service. The work is part of the Ph.D. project of E. Bahn who would like to thank the Ecole Doctorale de Physique of the Universit\'{e} de Grenoble for funding. The authors acknowledge use of and support by the Cambridge Atom Scattering Facility (\url{https://atomscattering.phy.cam.ac.uk}) and EPSRC award EP/T00634X/1.

\subsection*{Author contributions} 
A.T. and E.B. performed the experimental measurements and carried most of the data analysis out. J.Z. and D.W. provided help with sample preparation and measurements. M.S. performed the vdW corrected DFT calculations. A.T. ran and interpreted the kinetic MC simulations with J.Z. being involved in the development of the MC code. A.T., E.B., A.P.J., J.E. and W.A. developed the physical interpretation of the data. P.F. and S.J.J. contributed to the conception of the project and all authors discussed the results and contributed to writing the manuscript.

\subsection*{Supplementary Information} 
Additional supplementary information accompanies the paper.

\pagebreak

\begin{bibunit}
\setcounter{equation}{0}
\setcounter{figure}{0}
\setcounter{table}{0}
\setcounter{section}{0}
\makeatletter
\titleformat*{\section}{\large\bfseries}

\renewcommand{\thefigure}{\textbf{\arabic{figure}}}
\renewcommand{\thetable}{\arabic{table}}
\renewcommand{\figurename}{\textbf{Supplementary Fig.}}
\renewcommand{\tablename}{Supplementary Table}
\renewcommand{\refname}{\large Supplementary References}
\renewcommand{\figureautorefname}{Supplementary Figure}
\newcommand{\subfigureautorefname}{Supplementary Figure}
\renewcommand{\tableautorefname}{Supplementary Table}

\twocolumn[\begin{@twocolumnfalse}
\textbf{\Large{{Supplementary Information\\Motion of water monomers reveals a kinetic barrier to ice nucleation on graphene.\\}}
}
~\\
\end{@twocolumnfalse}]
\setcounter{page}{1}

\section{Supplementary experimental details}
\subsection{Sample preparation}
\label{S:prep}
The growth and characterisation of the graphene layer on a Ni(111) surface has been published  elsewhere\cite{tamtogl2015}. In short, the nickel (111) single crystal used in the study was mounted onto a sample holder and can be heated radiatively by a filament on the back of the crystal, or cooled to 100\,K using liquid nitrogen. Prior to the measurements, the Ni surface was cleaned by repeated cycles of Ar$^{+}$ sputtering and annealing at 870\,K. A monolayer of graphene on Ni(111) was grown by dosing ethene (C$_2$H$_4$) while holding the crystal at 730\,K for several hours.

The sample temperature was measured using a chromel-alumel thermocouple. While absolute temperatures can be determined to an accuracy of $\pm 5\,\mbox{K}$, relative temperature values were determined with an accuracy of $\pm 0.1\,\mbox{K}$, which was also confirmed by the reproducibility of the adsorption and dynamics measurements.

Water adsorption and desorption processes were studied during dosing with a precise water pressure control obtained by a motorised leak valve attached to the dosing supply. The leak valve itself was regulated by a feedback control system in order to maintain a constant pressure. Adsorption was measured at sample temperatures of $100$, $110$, $125$, $130$ and $150\,\mbox{K}$ at a typical dosing pressure at the surface of $(2-20)\cdot 10^{-9}~\mbox{mbar}$.

The helium reflectivity, at 100\,K as shown in Figure \textbf{3} of the main article decreases continuously and remains low when the dose is stopped. Such a behaviour is typical of  disordered structures forming, e.g. the growth of an amorphous layer. The formation of amorphous ice layers on surfaces, commonly referred to as amorphous solid water (ASW) has been observed since the 1960s\cite{mcmillan1965}. For example, recent isothermal desorption measurements of water on highly oriented pyrolytic graphite (HOPG) at 100\,K, showed a glass transition accompanied by a change in desorption rate and a growth of 3D water islands, rather than a wetting of the graphite surface\cite{lofgren2003}. At 110\,K and 125\,K, the helium reflectivity also decays during deposition, but reaches saturation. Based on the fact that exactly the same diffraction pattern is observed as from clean graphene, the deposition has been interpreted as forming separated islands of ice, leaving areas of bare graphene in between (see main text). 

For sample temperatures above 120\,K, when the applied pressure is reduced after dosing the helium reflectivity recovers very quickly. The system is thus in a dynamic equilibrium where small changes in the pressure immediately affect the coverage and hence the reflectivity. While with increasing overpressure the coverage increases, with increasing surface temperature the dynamic equilibrium is also reached faster. Within the available temperature range -- where we could observe diffusion and where we are able to obtain a constant coverage by applying an overpressure -- it was found that measurements at 125\,K provided the best trade-off in order to clearly see dynamics and maintain a constant coverage. Dephasing rate measurements at 125\,K are reported in the main text (figure \ref{fig:Fig1}b), but further measurements at 130\,K, indicate that within experimental uncertainties the measurements have the same variation with scattering momentum transfer, $\Delta K$, and thus the same mechanism of motion at different temperatures. Finally, at even higher surface temperatures, i. e. around 150\,K and above, negligible adsorption is observed, even when pressures of up to $1.5 \cdot 10^{-7}~\mbox{mbar}$ were applied.

\subsection{Coverage calibration and monomer lifetime}
\label{S:Coverage}
All measurements were performed at the same coverage of $0.07\,\mbox{ML}$, which we define using a particular value of reflectivity, $I/I_0 = 0.25$. The reflectivity is adjusted, at each temperature, by varying the overpressure of water vapour from the capillary doser (see \nameref{methods}). We estimate the value of the coverage when $I/I_0 = 0.25$ using the measured dynamical data (figure \ref{fig:Fig5}, main paper) and, in particular, the shape and position of the features arising from the pairwise inter-adsorbate forces. The method provides an absolute measure of coverage since the prominent minimum in the data ($|\Delta \mathbf{K}| \approx 0.8\,\mbox{\AA}^{-1}$) corresponds to a quasi-hexagonal overlayer with a spacing that defines the coverage. Varying the coverage in the kinetic Monte-Carlo calculations leads to the red curve in figure \ref{fig:Fig5}, which corresponds to a coverage of $(0.07 \pm 0.02)\,\mbox{ML}$ (where
1 ML is defined as one water molecule per adsorption site).
\begin{figure}[htbp]
     \centering
     \includegraphics[width=0.82\linewidth]{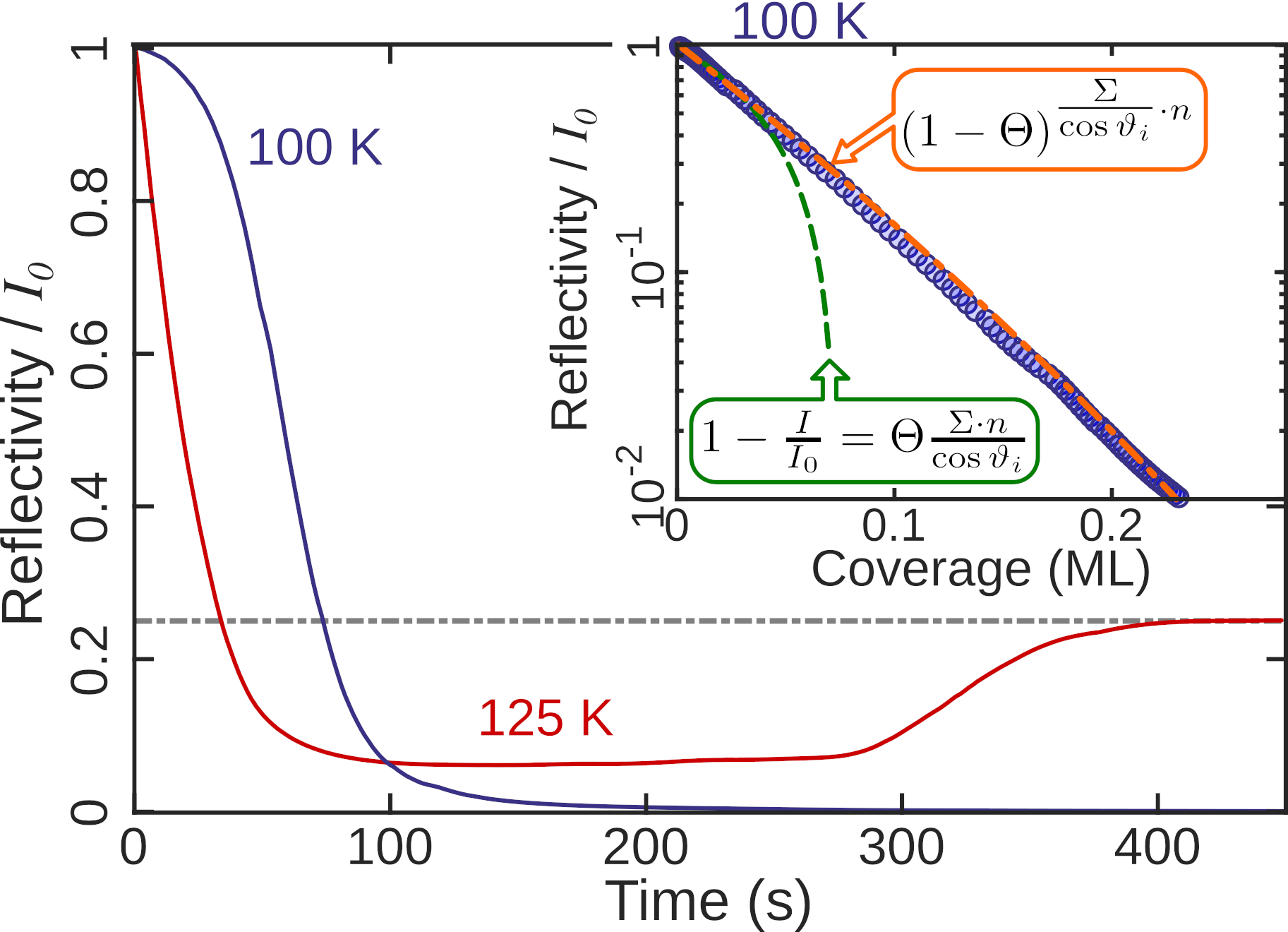}
     \caption{\textbf{Helium reflectivity during water deposition.} The figure shows the time dependence of the reflectivity at 125\,K (red curve) and 100\,K (blue curve). The blue points in the inset show the helium reflectivity during water uptake at 100\,K, converted to a coverage scale based on integrating exposure and assuming perfect sticking, as described in the text. The reflectivity follows the so-called lattice gas formula\cite{poelsema1989} (orange dash-dotted line), which assumes perfectly diffuse scattering from randomly distributed adsorbates on lattice sites, allowing for a cross-section overlap.}
     \label{fig:S1}
\end{figure}

A second method of coverage calibration is less direct but it serves to confirm the internal consistency in our measurements and their interpretation. Here, we use the dosing rate to calculate the coverage, assuming a constant sticking coefficient of unity (see Refs.\cite{chakarov1995b,chakarov1995,bolina2005}, with monolayer coverage corresponding to $n=0.115$ molecules$/\mbox{\AA}^2$,\cite{smith2014} which is close to the density of an ice \Romannum{1}\textsubscript{h} overlayer\cite{chakarov1995}). The rate of impingement is calculated from the measured partial-pressure of water in the chamber and the known enhancement generated by the microcapillary array\cite{ward2013}. Using that approach we can show that upon water adsorption at 100\,K, the reflectivity follows a model for random adsorption (stick and sit, see Ref. \cite{poelsema1989}), with remarkable agreement over two-orders of magnitude (\autoref{fig:S1}). Hence it confirms our assumption that the water monomers are static at 100\,K. Also, the method allows us to estimate a scattering cross section for isolated water monomers of $\Sigma  = ( 120 \pm 20 )\,\mbox{\AA}^2$, consistent with the coverage calibration derived from the dynamic measurements. Here, most of the quoted tolerance arises from uncertainties in the doser calibration and doser position relative to the sample. The determined cross section is in good agreement with other water cross sections in the literature on similar systems (e.g. $\Sigma = 130\,\mbox{\AA}^2$ in Ref \cite{glebov1999}).

The fraction of the surface covered by ice islands cannot be estimated from the scattering data since the diffraction patterns with and without water adsorption are essentially identical (figure \ref{fig:Fig2}d, main paper). It follows that the area of the surface covered by ice is extremely small. We can estimate the separation of ice islands from the r.m.s. distance travelled by an H$_2$O monomer using known conditions for dosing and the measured diffusion constant. At 125\,K the impinging flux of molecules was $2.5\cdot 10^{17}\,\mbox{m}^{-2}\mbox{s}^{-1} $ and the corresponding rate of adsorption is $0.022$ monolayer per second. Taking the coverage of water monomers, at equilibrium, to be $0.07\,\mbox{ML}$ we obtain $3.4\,\mbox{s}$ as the average lifetime of an H$_2$O monomer, after adsorption and before being bound to an island. From the dynamics measurements in the main part of the manuscript we know that the hopping rate at 125 K is $1.5\cdot 10^{10}~\mbox{s} ^{-1}$. The mean jump length during diffusion is $3.3\,\mbox{\AA}$ and it follows that a monomer travels about $70\,\upmu\mbox{m}$ before sticking to an island. The island separation must therefore be on a similar scale, which leads to the approximate scale-bar indicated in figure \ref{fig:Fig2}c (main paper).

\subsection{Supplementary desorption measurements}
\label{S:TDS}
Several groups have conducted thermal desorption spectroscopy (TDS) measurements of water on the (0001) basal plane of graphite. Consistently, a single desorption peak was observed that corresponds to a desorption energy in the range of $0.4$ to $0.5$\,eV which is close to the sublimation enthalpy of ice, $0.49~\mbox{eV}$\cite{chakarov1995,ulbricht2006,bolina2005}. The observed desorption energy does not change with coverage, indicating the formation of separated islands on the graphene surface\cite{ulbricht2006}. 

On the surfaces of graphene/Ni(111) and of graphene/Ir(111), TDS spectra reveal pseudo-zeroth order desorption and desorption energies of $(356\pm23)\,\mbox{meV}$ in the first case, and $(585\pm31)\,\mbox{meV}$ in the latter case, respectively, were found\cite{bottcher2011}. Smith \emph{et al.} report a desorption energy of about $490~\mbox{meV}$ for graphene/Pt(111), however they noted that the desorption of the water monolayer is complicated by de-wetting upon desorption and only multi-layer water films show zero-order desorption\cite{smith2014}.

\begin{figure}[htb]
     \centering
     \includegraphics[width=0.8\linewidth]{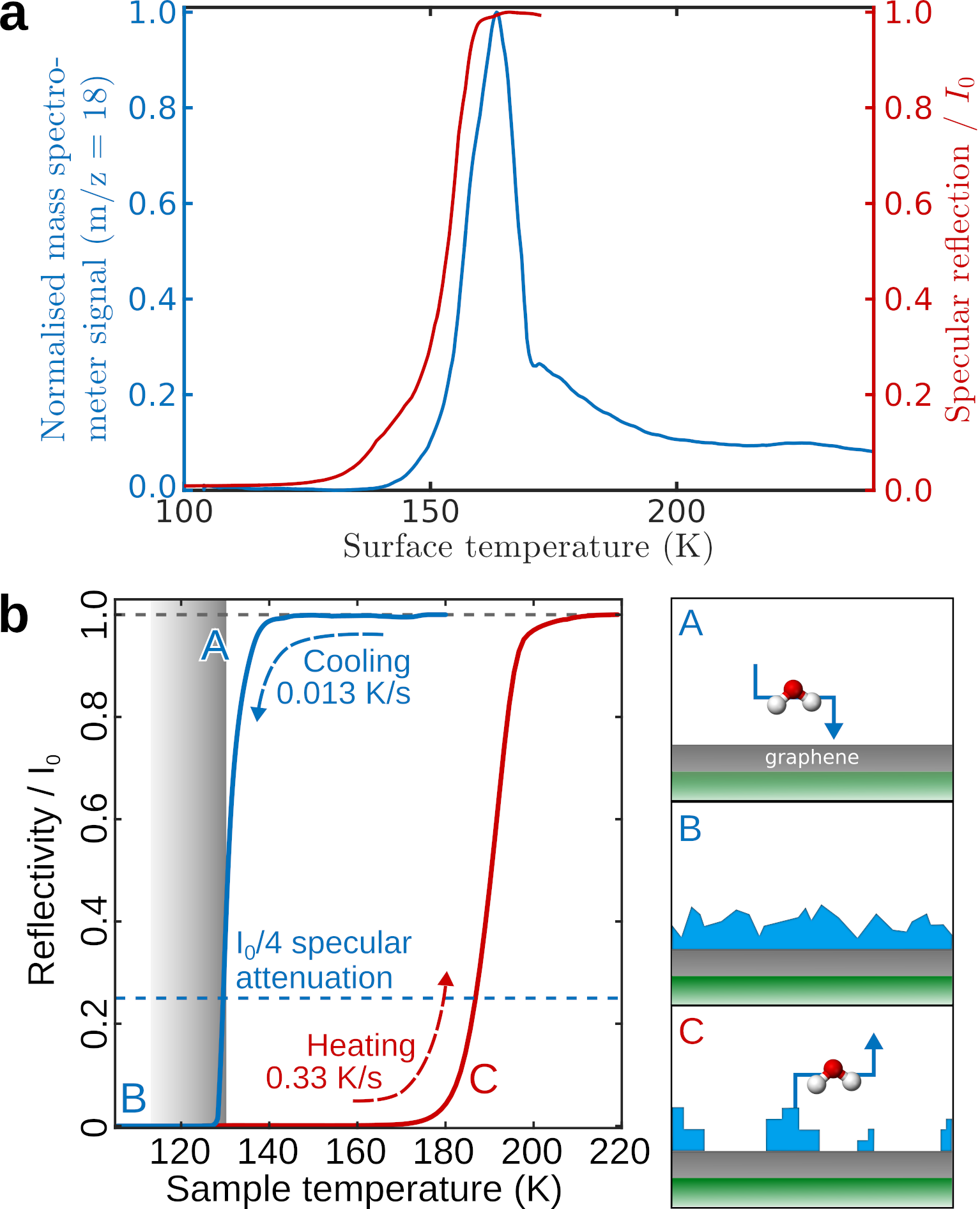}
     \caption{\textbf{Thermal desorption and isobaric adsorption of thick films. a} Thermal desorption spectroscopy (left axis) together with the specular He reflection (right axis) for thick films of water deposited on graphene, prior to heating the sample. \textbf{b} Isobaric adsorption for a partial H$_2$O pressure of $1.4\cdot 10^{-8}~\mbox{mbar}$, showing the variation of the specular He reflection as a function of the surface temperature $T_S$. Starting from the top right corner ($T_S = 180~\mbox{K}$), the sample is cooled down to 100 K and then heated up again. The signal follows a hysteresis, with desorption occurring at a higher temperature than adsorption, caused by the nucleation kinetics on the surface which is illustrated on the right-hand side. The shaded temperature region represents the temperature window where dynamics is observable.}
     \label{fig:S2}
\end{figure}

We have also conducted thermal desorption spectroscopy while monitoring the $m/z = 18$ peak on a mass spectrometer and simultaneously measuring the helium reflectivity (\autoref{fig:S2}a). A single desorption peak with a maximum at 163\,K coincides with a rapid recovery of the specular signal. The Redhead equation can be applied, in order to estimate the desorption energy $E_d$. Using $\nu = 9\cdot 10^{-14}~\mbox{s}^{-1}$ according to Ulbricht \emph{et al.}\cite{ulbricht2006} for the peak maximum at $(163\pm5)\,\mbox{K}$ at a heating rate $\beta = 0.22~\mbox{K}\cdot \mbox{s}^{-1}$, we obtain a desorption energy of $E_d = (520 \pm 20)\,\mbox{meV}$.

Furthermore, we can use the recovery of the helium reflectivity to determine the desorption energy. We exposed the graphene surface to a water overpressure of $2\cdot 10^{-8}~\mbox{mbar}$ and waited until the system was in equilibrium, before turning off the exposure and monitoring the reflectivity recovery. From this we calculated the corresponding surface coverage as a function of time. The surface coverage first rises during exposure and then decays exponentially after exposure has been turned off. The initial desorption rate, which is identical to the exponential decay rate, exhibits an activated temperature dependence. The desorption energy can then be determined from the slope in an Arrhenius plot and gives a value of $E_d = (510 \pm 10)\,\mbox{meV}$, in good agreement with the conventional TDS method.

As mentioned in the main text, it suggest that water molecules tend to desorb from the surface of water islands into the vacuum rather than as individual molecules which are adsorbed on the bare graphene surface - with the latter being more likely to diffuse and bind to an island. These findings are further supported by the diffraction measurements stated in the main text and the mentioned de-wetting upon desorption as observed for water on graphene/Pt(111)\cite{smith2014}.
\begin{table*}[!htp]
\caption{Comparison of experimentally determined diffusion parameters for water monomers on different substrates, including the activation energy, $E_a$, the diffusion constant, $D_0$, in Arrhenius pre-exponential form, and the hopping attempt rate, $\Gamma_0$.}
\centering
\begin{tabular}{l c . c c c c}
\toprule
Substrate & Adsorbate  & \multicolumn{1}{c}{$E_a$ (meV)} & \multicolumn{1}{c}{$D_0$ (m$^2$s$^{-1}$) } &  \multicolumn{1}{c}{$\Gamma_0$ (s$^{-1}$)} & \multicolumn{1}{c}{$T$ range (K) }  & Ref. $~$ \\ 
\midrule
\textbf{Graphene/Ni(111)} & \textbf{H$_2$O} & \textbf{60} & $\mathbf{1.1\cdot 10^{-7}} $ & $\mathbf{4.0\cdot 10^{12}}$ &  $\mathbf{113-130}$ & \textbf{present work} \\
Cu(111) & D$_2$O  & 75 & $1.8\cdot 10^{-8}$ & $1.8\cdot 10^{11}$ & $23-29$ & \cite{bertram2019} \\
Bi$_2$Te$_3$(111) & H$_2$O & 34 & $1.3\cdot 10^{-8} $  & $1.7\cdot 10^{11}$ & $130-160$ & \cite{tamtogl2020} \\
NaCl(001)/Ag(111) & D$_2$O & 149 & $1.5\cdot 10^{-8}$ & $1.0\cdot 10^{12}$  & $~42-~52$ & \cite{heidorn2015} \\
\bottomrule
\end{tabular}
\label{tab:CompTab}
\end{table*}

\subsection{Isobaric adsorption} 
\autoref{fig:S2}b shows an isobaric deposition curve of water on the graphene/Ni(111) surface: At a constant partial pressure of H$_2$O of $1.4\cdot 10^{-8}~\mbox{mbar}$ the temperature of the crystal is decreased from 180 K down to 100 K. There is no significant decrease in the intensity until the crystal reaches about 140 K where the intensity of the specular peak falls off sharply corresponding to the commencement of adsorption. The specular intensity drops to almost zero when the crystal temperature has reached 100 K. Upon starting to heat the system under the same conditions the specular intensity does not increase before we reach temperatures above 160 K. Hence we are observing a hysteresis, with desorption occurring at a higher temperature than adsorption.

One reason for this behaviour might be the higher heating rate - i.e. with increasing heating rate the desorption maximum shifts to higher temperatures. However, it cannot explain a shift of this magnitude. Instead the hysteresis shows that there is a kinetic barrier to nucleation on the surface. Upon cooling the system down, the intensity drop occurs at much lower temperatures (at about 140 K) because adsorption on the hydrophobic bare graphene surface\cite{aria2016,zhao2017} is less likely than onto an existing island of ice. Ice growth on the graphene surface is delayed because some clustering centres on the surface are necessary to allow the process to start\cite{bottcher2011}. On the other hand, upon heating, the surface is already covered with amorphous ice, from which it is harder to remove a molecule and hence the intensity only starts to recover at about 160 K.

\section{DFT calculations}
\label{S:DFT}
\begin{figure}[htbp]
     \centering
     \includegraphics[width=0.95\linewidth]{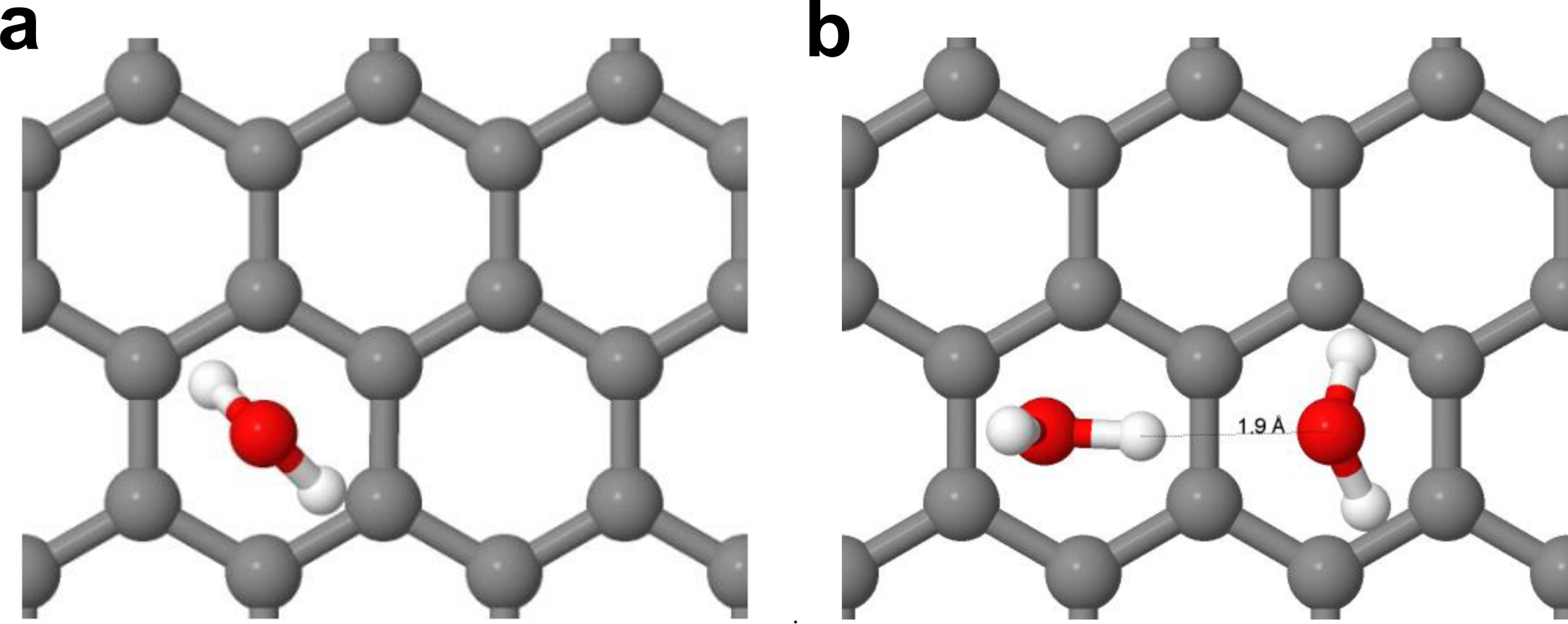}
     \caption{\textbf{Adsorption geometries of water according to DFT calculations. a}  Adsorption geometry for a single H$_2$O molecule on graphene according to our DFT calculations. \textbf{b} Optimised adsorption geometry after the formation of an H$_2$O dimer on graphene.}
     \label{fig:S3}
\end{figure}

The density functional theory (DFT) approach has been applied a number of times to the adsorption of water on graphene. DFT calculations generally agree that the potential energy surface is rather flat and that the binding energy depends more on the orientation than on the position of the adsorbent. Most calculations predict a preferential water adsorption with the hydrogen atoms pointing downwards (see figure \ref{fig:Fig4}b and \autoref{fig:S3}a). We obtained an adsorption energy, $E_{ads}$, of about $250\,\mbox{meV}$ for the global minimum, which is located at the centre of the graphene hexagonal unit cell. The value is in very good agreement with the 183\,meV obtained by Li \textit{et al.}\cite{li2012}, while other DFT calculations give slightly smaller adsorption energies\cite{bottcher2011,ma2011,leenaerts2009,ambrosetti2011,freitas2011}. We also find the water molecule is adsorbed with the orientation usually predicted, i.e. that it has the two OH bonds pointing towards the surface, so that the plane of the molecule is perpendicular to the surface plane itself. A general agreement on an adsorption distance of about $3.3\,\mbox{\AA}$ can be observed between our results and previous reports\cite{leenaerts2009,wehling2008,hamada2012,li2012}.

In addition to calculations on bare graphene, we performed a set of vdW DFT calculations including a Ni substrate (modelled as a five-layers nickel slab with a $(\sqrt{7}\times \sqrt{7} )$ surface unit cell) to verify that the Ni substrate does not affect significantly the nature of water interaction with graphene. We found that the water to graphene distance and the orientation of the water monomers are very close to the calculations on the pristine graphene without the Ni substrate, although the absolute adsorption energy with the substrate is slightly different, $\approx230\,\mbox{meV}$ compared to $\ 250\,\mbox{meV}$ for suspended graphene. Including the substrate necessarily reduces the size of the surface unit cell and, by omitting the Ni, it is possible to increase the unit cell significantly (a 9-fold increase in the number of carbon atoms). The larger supercell allows us to model adsorption at significantly lower coverages, much close to those in the experiment. For these reasons, we used a larger graphene supercell without substrate to investigate the dynamics of water motion at low coverage, as presented in the manuscript.

We estimate the magnitude of the energy barrier to dimer formation by performing a series of DFT calculations for two water monomers (in the same orientation as the isolated monomer) with varying distance. The resulting barrier is about 90\,meV, while the binding energy of the dimer is approximately $200\,\mbox{meV}$. Thus, the barrier to dimer formation is significant and we can conclude that, once a dimer forms, it will rarely dissociate. The dimer exhibits then a total adsorption energy of $595-696\,\mbox{meV}$ (depending if the substrate is frozen or not). These results support also our observation of hysteresis during isobaric cooling/heating and they are consistent with previous calculations of of H$_2$O clusters adsorbed on graphite where the association energy (including re-orientation, binding and adsorption) with the cluster is in the range of $450-500$ meV, while the binding energy of a monomer to the graphene surface is much lower\cite{leenaerts2009,wehling2008}.

In the measured $\Delta K$-range of this study, the dephasing rate $\alpha$ is in the order of $10~\mbox{ns}^{-1}$ at 125 K. This temperature corresponds to a mean kinetic energy in the order of 10 meV. Since the adsorption energy of an H$_2$O molecule in an ice cluster is predicted to be in the order of 500 meV\cite{leenaerts2009}, while for the adsorption energy of a molecule on the graphene surface, values in the order of $100-200$ meV have been calculated, one would expect to observe the diffusion of H$_2$O on graphene, rather than on the surface of an ice cluster. Together with the adsorption and diffraction results this is another evidence that we are seeing the motion of single water molecules on graphene.

\section{Supplementary discussion}
\label{S:Discussion}
\autoref{tab:CompTab} compares our activation energies and diffusion constants (top row) with water measurements on other substrates in the literature.

In addition, the diffusion of water on graphene has been recently studied by means of molecular dynamics (MD) simulations. Tocci \textit{et al.} predict a substantially lower macroscopic friction coefficient in comparison to adsorption on a hexagonal boron nitride surface\cite{tocci2014} and Park \textit{et al.} predicted fast diffusion with a diffusion constant $D = 2.6 \cdot 10^{-8}~\mbox{m}^2/\mbox{s}$\cite{park2010}. MD simulations of water nanodroplets on freestanding graphene\cite{ma2015} revealed a diffusion constant between $2 \cdot 10^{-7}~\mbox{m}^2/\mbox{s}$ and $8.6 \cdot 10^{-7}~\mbox{m}^2/\mbox{s}$ depending on the size of the droplet (at 298 K). Both values are way beyond the diffusion constant found in our experiments, yet they are considering the motion of water clusters and droplets at much higher temperatures (room temperature) rather than the diffusion of monomers.\\
The diffusion coefficient for single water molecules on graphene has been estimated to be  $6 \cdot 10^{-9}~\mbox{m}^2/\mbox{s}$ at a temperature of 100 K by Ma \textit{et al.} \cite{ma2011} which is somewhat closer to the conditions in our experiments. Indeed their value is closer to our result but still one order of magnitude larger. However, all calculations mentioned above were performed on free standing graphene while our measurements are on graphene/Ni(111) where the motion of the ripples which gives rise to the ultra-fast diffusion\cite{ma2015} is suppressed\cite{tamtogl2015}. Compared to the diffusion in (bulk) amorphous ice on the other hand, where for the translational motion $D_0$  is in the range of $(0.5-5)\cdot10^{-17}\,\mbox{m}^2/\mbox{s}$\cite{perakis2017}, or the diffusion of ASW at the liquid/ice interface with $D \approx 10^{-21}\,\mbox{m}^2/\mbox{s}$ at 125\,K\cite{xu2016}, the diffusion of water monomers on graphene with $D = 4.1\cdot10^{-10}\,\mbox{m}^2/\mbox{s}$ at 125\,K is incomparably faster.

\section{Supplementary details about single particle and collective motion}
\label{S:MCSupp}
The trajectories of the molecules versus time resulting from the KMC simulations can be used to calculate the intermediate scattering function (ISF) which is also obtained in the experiment. From the KMC simulation both the coherent and the incoherent ISF can be calculated. The subtle difference between the coherent and incoherent ISF is the averaging procedure. While the coherent ISF is obtained by averaging over all particles, the incoherent ISF is obtained by first calculating the ISF of a single particle followed by averaging over all particles. Details on how to obtain both the coherent and the incoherent ISF can be found elsewhere\cite{ward2020}.\\
\begin{equation}
\begin{aligned} 
I(\Delta K,t)_{coh} &= \frac{1}{N}  \sum_{j,k}^N \left\langle  \mathrm{e}^{  \mathbbm{i} \Delta \mathbf{K} \left( \mathbf{R}_j(t) - \mathbf{R}_k(0) \right) } \right\rangle \, , \\
I(\Delta K,t)_{inc} &= \frac{1}{N}  \sum_{j}^N \left\langle  \mathrm{e}^{  \mathbbm{i} \Delta \mathbf{K} \left( \mathbf{R}_j(t) - \mathbf{R}_j(0) \right) } \right\rangle \, \, , 
\label{eq:ISF}
\end{aligned}
\end{equation}
The ISFs obtained from the simulation are then analysed in the same way as the experimental data: The ISF is fitted with a single exponential decay which allows to determine the dephasing rate $\alpha (\Delta K)$ from the simulation in analogy to the curve determined from the experiments. The trajectories from the KMC simulation can be used to calculate both the coherent and the incoherent ISF.

On the other hand, He spin-echo is a coherent scattering method, hence the measurements provide the coherent ISF. As shown for neutron scattering\cite{pusey1975} as well as for X-ray photocorrelation spectroscopy\cite{leitner2009} one can de-correlate  the effect of adsorbate interactions, i.e. obtain the corresponding incoherent $\alpha_{inc} (\Delta K)$ from the measured coherent $\alpha_{coh} (\Delta K)$\cite{ward2020}.\\
Following the derivation of Sinha and Ross\cite{sinha1988}, where the interaction forces are considered as a mean field, the scattering function $S_s$ of the non-interacting system becomes: 
\begin{equation}
S_s(\Delta K,\omega)=\frac{\Gamma_s(\Delta K ) \; c \, (1-c) }{\pi \, (\Gamma_s(\Delta K )^2 + \omega^2)},
\end{equation}
where $c$ is the concentration of dynamic adsorbates, and the quasi-elastic broadening $\Gamma_s$ follows the well established Chudley-Eliott lineshape (the analytical model as in the main text)\cite{ward2020}.

The only region where this approach does not apply is in the vicinity of the substrate diffraction peaks. Here the structure factor of the substrate becomes important while at the same time the uncertainty of the quasi-elastic amplitude becomes large. As a consequence the blue dots in Figure \ref{fig:Fig1}b of the main text show an offset for $\Delta K$ close to zero and around the diffraction peaks.

\begin{figure}[htb]
     \centering
     \includegraphics[width=0.8\linewidth]{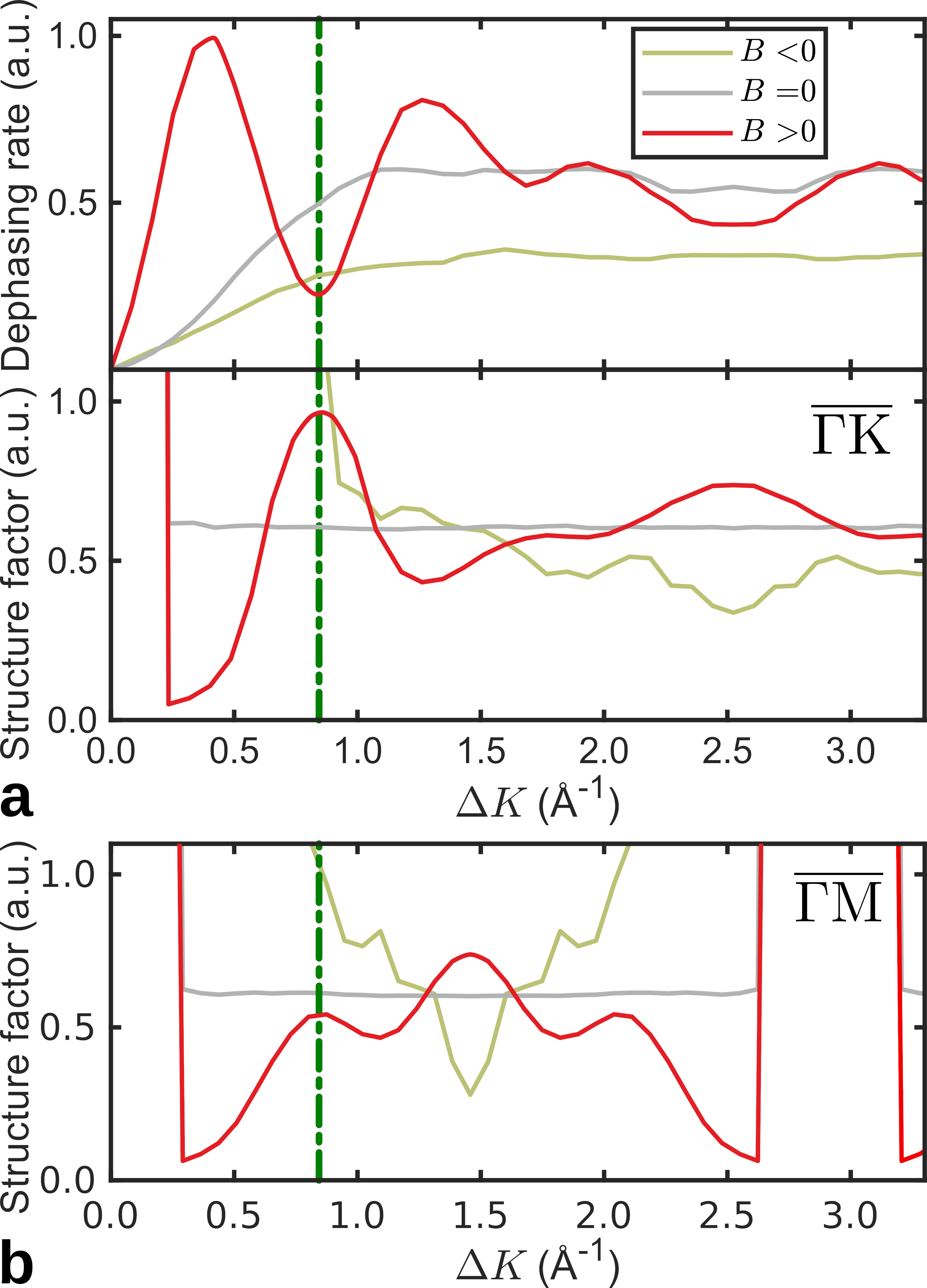}
     \caption{\textbf{a} Comparison between the dephasing rate $\alpha$ (upper panel) and the corresponding static structure factor $S (\Delta K)$ (lower panel) along the \GK -azimuth, as extracted from the KMC simulations at the same conditions (including temperature and water coverage) as for the experiment. Only repulsive interactions ($B > 0$) give rise to a clear peak in $S (\Delta K)$ at the same position where $\alpha ( \Delta K )$ shows a dip as illustrated by vertical dash-dotted line. \textbf{b} Static structure factor $S (\Delta K)$ along the \GM -azimuth. }
     \label{fig:S6}
\end{figure}

Finally, here we note again that only the implementation of repulsive interactions in the KMC simulation can reproduce the peak-and-dip structure as evident in the experimental data. \autoref{fig:S6}a shows both the dephasing rate $\alpha$ (top panel) and the corresponding static structure factor $S (\Delta K)$ (lower panel) as extracted for the KMC simulations along the \GK -azimuth. We see that only for repulsive forces ($B > 0$, \autoref{eq:MCdipole} in the main text) there appears a clear peak in $S (\Delta K)$ at the same position, where $\alpha ( \Delta K )$ shows a dip, as illustrated by the dash-dotted vertical line. 

The repulsive forces between the adsorbates give rise to a deviation of the dephasing rate $\alpha(\Delta K)$ with respect to the analytical expression (\autoref{eq:CE} in the main text). Adsorbates repelling each other prefer a long-range quasi-hexagonal structure leading to a preferred, coverage dependent average distance (see \ref{S:Coverage}) between the adsorbates and reduced mobility on these length scales\cite{jardine2009,alexandrowicz2006,nissila2002}. At the same time, when adsorbates approach each other their mobility increases compared to the non-repelling case. The result is a peak at lower $\Delta K$ followed by a dip feature, termed as ``de Gennes narrowing'' as illustrated by the red line in the top panel of \autoref{fig:S6}a. We see that the grey line from the KMC simulation without repulsive interactions follows the same sinusoidal curve as for the analytical expression while the red line - illustrating the case for inter-adsorbate repulsion - exhibits a peak appearing at low $\Delta K$ values due to the increased mobility at certain length scales, followed by a dip (vertical dash-dotted line) occuring at the length scale of the quasi-hexagonal arrangement. The location of this dip corresponds to a peak in the static structure factor\cite{serra2002}, as seen in the lower panel of \autoref{fig:S6}a and \autoref{fig:S6}b.

We should also note that the current observation of long-range repulsive interactions does not exclude the possibility of short-range attractive interactions and it is the implementation in the KMC that reproduces the feature in the experimental data. Short-range interactions may rather occur within a length scale that corresponds to intra-cell diffusion \cite{stradner2004} while the discrete grid in terms of the KMC simulations allows just for interactions at the the inter-cell diffusion length-scale to be taken care of. 

\setlength{\bibsep}{0.0pt}
\putbib[literature]
\end{bibunit}

\end{document}